\documentclass[journal=jctcce,manuscript=article]{achemso}

\usepackage{soul}
\usepackage{xcolor}
\usepackage{multirow}
\usepackage{arydshln}
\usepackage{gensymb}
\usepackage{graphicx}            
\graphicspath{ {images/} }

\author{Santosh Adhikari}
\email{tuf60388@temple.edu}
\phone{+1 267 909 1660}
\author{Hong Tang}
\author{Bimal Neupane}
\affiliation[Temple University]
{Department of Physics, Temple University, Philadelphia, PA, 19122}
\author{G\'abor I. Csonka}
\affiliation[Department of Inorganic and Analytical Chemistry]
{Department of Chemistry, Budapest University of Technology and Economics, Budapest, Hungary, H-1111}
\author{Adrienn Ruzsinszky}
\affiliation[Temple University]
{Department of Physics, Temple University, Philadelphia, PA, 19122}
\email{tuf27796@temple.edu}
\title[An \textsf{achemso} demo]
  {Molecule-surface interaction from van der Waals-corrected semilocal density functionals: the example of thiophene on transition-metal surfaces}

\begin{document}
\begin{abstract}
  Semi-local density functional approximations are widely used. None of them can capture the long-range van der Waals (vdW) attraction between separated subsystems, but they differ remarkably in the extent to which they capture intermediate-range vdW effects responsible for equilibrium bonds between neighboring small closed-shell subsystems. The local density approximation (LDA) often overestimates this effect, while the Perdew-Burke-Ernzerhof (PBE) generalized gradient approximation (GGA) underestimates it. The strongly-constrained and appropriately normed (SCAN) meta-GGA often estimates it well. All of these semi-local functionals require an additive non-local correction such as the revised Vydrov-Van Voorhis 2010 (rVV10) to capture the long-range part. This work reports adsorption energies and the corresponding geometry of the aromatic thiophene (C$_4$H$_4$S) bound to transition metal surfaces. The adsorption process requires a genuine interplay of covalent and weak binding and requires a simultaneously accurate description of surface and adsorption energies with the correct prediction of the adsorption site. All these quantities must come from well balanced short and long-range correlation effects for a universally applicable method for weak interactions with chemical accuracy.
\end{abstract}

\section{Introduction}
The development and assessment of various vdW methods has been an intensive area of research in the past decade. \cite{langbein1976van,parsegian2005van,ajstone1996,kaplan2006intermolecular,french2010long,jeziorski1994perturbation} Now the scientific community possesses a broad range of approximations$\cite{dion2004van,johnson2005post,tkatchenko2012accurate}$ of useful but limited accuracy. vdW methods approximate the long-range correlation which arises from the physics of collective plasmon oscillations. Wavefunction-based approximations such as Coupled Cluster (CC) methods naturally include vdW interactions, but are practically beyond the reach of the condensed matter community at this time. Alternatively the Random Phase Approximation (RPA) is nearly exact for the long-range, and is regarded as a benchmark to assess vdW methods in condensed matter physics.$\cite{dobson2005soft}$ Even with increased computational power and increasingly efficient implementations, RPA has a limited practicality for materials.$\cite{nguyen2010first}$\\
Most current vdW methods have been developed within the density functional context, in which the construction of self-consistent orbitals is a parallelizable step. Semilocal density functional theory (DFT) can be accurate for the short-range correlation, but misses the long-range part. The long-range vdW component is captured by either pairwise vdW methods$\cite{tkatchenko2009accurate,grimme2004accurate,becke2007exchange}$ or by nonlocal correlation functionals.$\cite{dion2004van,vydrov2010nonlocal,sabatini2013nonlocal}$ Each of these approximations is then added to an appropriately chosen semilocal exchange-correlation functional.  The vdW-DF$\cite{dion2004van}$ and VV10$\cite{vydrov2010nonlocal}$ non-local correlation functionals are based on approximations for the polarizability, and VV10 has a fitted short-range attenuation parameter ($\textit{b}$) that adapts to the semilocal term. Many of the current vdW models are reasonably accurate and efficiently applicable to geometries and binding energies.\\
The recent rVV10$\cite{sabatini2013nonlocal}$ correlation functional has been combined$\cite{peng2016versatile}$ with the  SCAN$\cite{sun2015strongly}$ meta-GGA and has been successfully applied to various systems including interlayer binding energies, adsorption energies and structural properties.$\cite{sun2016j}$ One major advantage of this method is its computational efficiency. Although SCAN+rVV10 delivers a generally reasonable description for various properties, it gives a disappointing treatment for some others. Examples include the overstructured radial distribution functions in liquid water,$\cite{wiktor2017note}$ inaccurate structural and mechanical properties in PPTA,$\cite{Yu2019}$ and inaccurate prediction of ground state properties of MnO and CoO.$\cite{peng2017synergy}$
While SCAN captures intermediate-range vdW interactions, it may capture too much.  revSCAN,$\cite{mezei2018simple}$ a revised version of SCAN was constructed to diminish the intermediate-range vdW interaction.\\
This work explores the accuracy and precision of the SCAN+rVV10 and the revSCAN+rVV10 approximations for thiophene adsorption on the surface of coinage metals. For comparison we also include several GGA-based semilocal exchange-correlation functionals with rVVV10 correction into this assessment.\\
The adsorption of molecular species on metal surfaces is a relevant problem for both computational simulations and industry. In general the adsorption of organic species on metal surfaces can be a synergy of chemi - and physisorption, however, recent works on the adsorption of benzene on the surface of coinage metals reveal the large role of weak interactions.$\cite{liu2015quantitative}$ A recent work reported accurate SCAN+rVV10 binding energies for the adsorption of benzene on transition metal surfaces.$\cite{peng2016versatile}$ A universally accurate approximation can be expected to capture adsorption sites, surface and binding energies simultaneously in the adsorption process.\\
The thiophene molecule is the smallest aromatic sulfur-containing compound. It is a natural choice as a test case for simulations. Thiophene is also a good test to study reactions that follow the catalytic desulfurization on metal or semiconducting surfaces. The adsorption of the thiophene molecule on metal surfaces turns out different than the extensively studied case of benzene adsorption. Depending on the underlying metal, the adsorption of thiophene on the metal surface can show some chemisorption character$\cite{callsen2012semiempirical}$ whose description requires a very accurate balance of local and nonlocal correlation in the meta-GGA and its partner van der Waals approximation.$\cite{tonigold2010adsorption}$
A simultaneously correct description of the adsorption energies and sites is a challenge to density functional theory.$\cite{sun2011improved}$\\~\\
\textbf{\Large{Meta-GGA density functional approximations}}\\
Exchange-correlation approximations within DFT can be broadly categorized by the different ingredients of the 5 rungs of a ladder of increasing nonlocality.$\cite{perdew2001jacob}$ The accuracy of an approximation usually increases when more ingredients are included. The increased accuracy often comes at the price of deteriorated efficiency, especially when non-local information is included.\\
Among the most accurate density functional approximations are the meta-generalized approximations (meta-GGA’s). Meta-GGA’s constitute the third-rung of the ladder. Commonly used meta-GGAs include one more ingredient beyond the GGA level, the kinetic energy density $\tau(r)=\frac{1}{2} \sum_{i=0}^{occ}|\nabla\phi_{i}|^{2}$ where the $\phi_{i}$'s are Kohn-Sham orbitals. Different dimensionless variables built from $\tau(r)$  have been proposed for use in meta-GGAs. The most successful dimensionless variable so far$\cite{sun2012communication,zhao2006new}$ is $\alpha(r)=\frac{\tau(r)-\tau^{vW} (r)}{\tau^{UEG}(r)}$, where 
$\tau(r)^{vW}= \frac{|\nabla n(r)|^{2}}{8n(r)}$ is the von-Weisz\"acker kinetic energy density of a single-orbital system and $\tau^{UEG} (r)=\frac{3}{10} {(3\pi^{2})}^{\frac{2}{3}}n(r)^{\frac{5}{3}}$ is the kinetic energy density of the uniform electron gas. The $\alpha(r)$  variable is an iso-orbital indicator, recognizing different types of orbital overlap environments and directly related to the electron localization function. Single-orbital regions are identified by $\alpha(r)=0$, highly-overlapped metallic regions with slowly-varying electron densities are revealed by $\alpha(r)\approx 1$, and regions of non-bonding density overlap by $\alpha(r)>>1$.\\
Madsen et al.$\cite{madsen2009treatment}$ showed that inclusion of the kinetic energy densities enables meta-GGA’s to distinguish between dispersive and covalent interactions.  A family of nonempirical constructions$\cite{sun2011improved}$ led to the development of the SCAN$\cite{sun2015strongly}$ meta-GGA. SCAN satisfies 17 exact constraints, while preserving the ability to capture intermediate-range weak interactions. With tests and assessments on diverse systems, the SCAN meta-GGA has been a success-story in the past three years.$\cite{sun2015strongly,sun2016j,shahi2018accurate,nepal2018rocksalt,nepal2019}$ \\
Through the $\alpha$ dependence of the interpolation functions for exchange and correlation energy, SCAN can capture intermediate-range dispersive interactions as shown in the following equations:
\begin{equation}
	f_{x}(\alpha)=e^{-c_{1x}\frac{\alpha}{1-\alpha}}.\theta(1-\alpha)-d_{x}e^{\frac{C_{2x}}{1-\alpha}}.\theta(\alpha-1)
\end{equation}
\begin{equation}
	f_{c}(\alpha)= e^{-c_{1c}\frac{\alpha}{1-\alpha}}.\theta(1-\alpha)-d_{c}e^{\frac{C_{2c}}{1-\alpha}}.\theta(\alpha-1)
\end{equation}
Many physical situations require the long-range part of the correlation, mathematically described by a double integral in the three-dimensional space, and not captured by any semilocal density functional. The vdW correlation functional by Vydrov and Van Voorhis (VV10)$\cite{vydrov2010nonlocal}$ and rVV10$\cite{sabatini2013nonlocal}$ are the examples for a long-range functional that allows the nonlocal correlation energy and its derivatives to be efficiently evaluated in a plane wave framework, as pioneered by Rom\'an-P\'erez and Soler.$\cite{roman2009efficient}$ The long-range correlation is a double integral:$\cite{vydrov2010nonlocal,sabatini2013nonlocal}$
\begin{equation}
	E_{c}^{nl}= \int dr n(r)[\frac{1}{2}\int dr'\phi(r,r')n(r')+ \beta]
\end{equation}
The VV10 and rVV10 corrections are designed to vanish for the uniform electron gas. This feature makes it possible to pair the nonlocal correlation energy with the semilocal exchange-correlation energy by utilizing a parameter to damp the intermediate and short range contribution of the former. A critical ingredient in the kernel is the local band gap, a quantity that accounts for density inhomogeneity and makes VV10 and rVV10 applicable for molecular systems.
Like VV10, the rVV10 correction has two adjustable parameters $\textit{C}$ and $\textit{b}$ inside the kernel $\phi$ that allow it to adopt to any semilocal functional.
The values of the $\textit{C}$ and $\textit{b}$ parameters for rVV10 for SCAN were 0.0093 and 15.7 respectively.\\
A universally applicable and accurate vdW approximation should benefit from the interplay of the nonlocal and semilocal functionals. Aside from the particular form of the vdW correlation functional, the choice of the exchange functional has received considerable attention within this work. The choice of semilocal exchange has already attracted interest in the context of GGA density functional. The revPBE-GGA exchange functional chosen for vdW-DF often leads to too-large intermolecular binding distances and inaccurate binding energies.$\cite{klimevs2009chemical}$ To remedy these problems, alternative "less repulsive" exchange functionals have been proposed, which when incorporated within the vdW-DF scheme lead to much improved accuracy.\\
Earlier attempts emphasized the improvement of vdW-DF by exploring and developing alternatives to the original revPBE exchange. These studies were limited to PBE-based GGA's, and the underlying semilocal exchange was fitted to the vdW functional in an empirical manner.\\~\\
\textbf{\Large{Benchmark binding energies for the adsorption of thiophene on metals}}\\
To properly assess the limitations of our approximations, we need accurate benchmark adsorption energies. After appropriate calibration, temperature programmed desorption (TPD) or thermal desorption spectroscopy can be used  to evaluate the activation energy of desorption. The binding energy might be estimated from the temperature of maximum desorption via Redhead’s analysis.$\cite{de1990thermal}$ However, the estimated binding energies might display an uncertainty larger than the chemical accuracy of 0.04 eV required for an accurate description of the adsorption.$\cite{christian2016surface}$ A considerably more accurate complete analysis method would lead to more accurate results,$\cite{de1990thermal,christian2016surface}$ but no such results are available for thiophene on coinage metal surfaces according to our knowledge. The nonlocal random phase approximation (RPA)$\cite{dobson2005soft,eshuis2011parameter}$ could be a reliable reference for long-range vdW interactions. RPA calculations are, however beyond reach for large supercells at this time. Here in this work we use an approximation that is robust enough and mimics the RPA binding energies almost perfectly for the interaction of graphene and metal surfaces. This approximation$\cite{tao2018modeling}$, which we will call PBE+vdW-dZK from this point onwards, models the long-range van der Waals correction for physisorption of graphene on metals with the damped Zaremba-Kohn (ZK)$\cite{lifshitz1956sov,zaremba1976van}$ second-order perturbation theory. In this model, quadrupole-surface interactions and screening effects are included.  This model relies on accurate static polarizabilities from higher-level calculations$\cite{kusantha2019,delaere2002influence}$ and predicts the vdW interaction from the $C_3$ and $C_5$ coefficients and the distances between the particle and surface plane through an expression whose large -z asymptote is
\begin{equation}
	E_{vdW} = [-\frac{C_3}{(z-z_{0})^3}-\frac{C_5}{(z-z_{0})^5}]f_d
\end{equation}
with z being the distance between the particle and the surface and $z_0$ the reference plane position.
The damping factor for eqn.(4) is
\begin{equation}
	f_d=\frac{x^5}{(1+gx^{2}+hx^{4}+x^{10})^{\frac{1}{2}}}  
\end{equation}
where, $x=\frac{z-z_{0}}{b} > 0$, $g=\frac{2 b^{2}C_3}{C_5}$ and $h=\frac{10 b^{4}{C_3}^{2}}{{C_5}^{2}}$. The cutoff parameter 'b' was choosen to be 3.3 bohr.$\cite{tang2018}$ 
Instead of taking the static dipole polarizability of the thiophene molecule, we base our $C_3$ coefficients on the "renormalized atom" approach.$\cite{tang2018}$ The best polarizability for a particular atom (H, C, or S) in thiophene is then renormalized as   
\begin{equation}
	\alpha_{renormalizedatom}=\frac{\alpha_{(freeatom)}}{4{\alpha_{(C)}}+4{\alpha_{(H)}}+{\alpha_{(S)}}} \alpha_{(thiophene)}
\end{equation}
With the static polarizabilities we can find separate $C_{3}$ and $C_{5}$ coefficients for each of the three elements in thiophene. The formula of renormalization that we are using is constructed for a "particle" interacting with a metal surface. A molecule such as thiophene, even if not of a large size, cannot be treated as a particle. Here we treat it as a collection of renormalized atoms. Treating the whole thiophene as a particle would make its quadrupole polarizability grow roughly as the 5/3 power of its dipole polarizability, and would overestimate $C_5$ significantly.\\~\\
\textbf{\Large{Parameterization of rVV10}}\\
In the present work we are following the approach of parameterization of rVV10 for SCAN.$\cite{peng2016versatile}$ Since the change of the value of $\textit{C}$ parameter does not significantly improve the binding curve,$\cite{sabatini2013nonlocal,peng2016versatile,peng2017rehabilitation}$ we keep the value of $\textit{C}$ fixed. But, we optimize the $\textit{b}$ value by fitting to the CCSD(T)$\cite{patkowski2005accurate}$ binding energy curve of the Argon dimer. Notice that a recent empirical potential energy function for Ar dimer$\cite{myatt2018new}$ showed excellent agreement with CCSD(T) and CCSDT(Q) results.$\cite{jager2009ab}$ The use of such ab initio derived potential functions for reference can be justified.$\cite{deiters2019two}$ For all the calculations, the Argon dimer was placed in a cubic supercell of 25 {\AA}. All the calculations were done using a single point gamma-centered k-mesh.\\
\begin{figure}[h!]
	
	\includegraphics[scale=0.32]{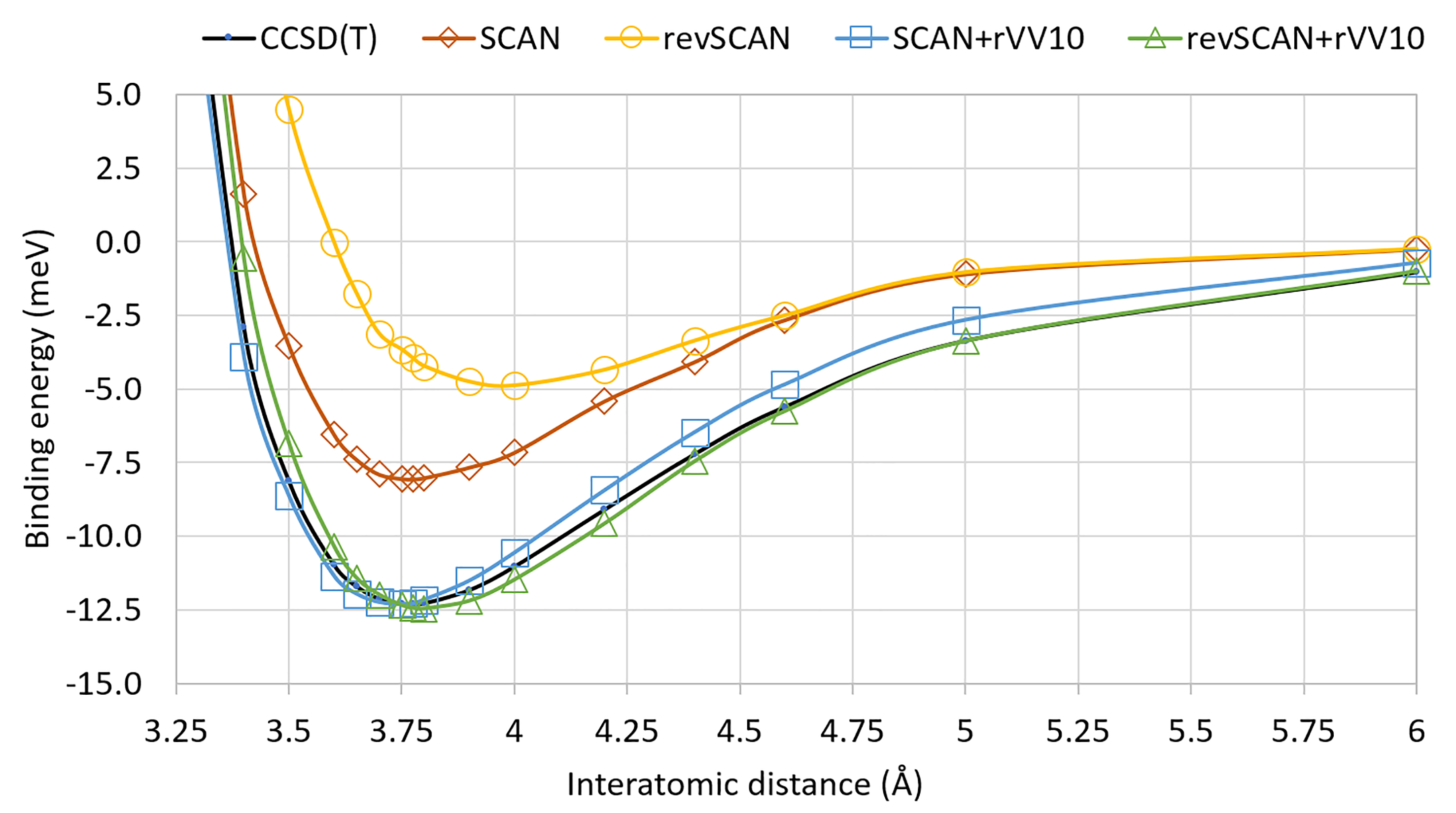}  
	\caption{The binding energy curve of the Ar dimer from SCAN, revSCAN and their corresponding rVV10 corrected versions with respect to CCSD(T) curve. The value of $\textit{b}$ for SCAN+rVV10 and revSCAN+rVV10  are 15.7 and 9.4, respectively.}
\end{figure}
Figure 1 displays the binding energy curves of the Argon dimer from SCAN and revSCAN and the corresponding rVV10 corrected version with the CCSD(T) data.$\cite{patkowski2005accurate}$ The revSCAN is more underbinding than SCAN in the intermediate range due to its construction, suggesting its need for a stronger van der Waals correction. We determine the $\textit{b}$ parameter for rVV10 for revSCAN by fitting to nine data points around the equilibrium distance with respect to the CCSD(T) binding curve. With 2.89\% of mean absolute percentage error (MAPE), the $\textit{b}$ parameter was determined to be 9.4. This value is slightly smaller than the $\textit{b}$ = 9.8 suggested in the original revSCAN+VV10. This smaller value leads to stronger dispersion interaction. The SCAN+rVV10 has MAPE of 3.32\%  with the original $\textit{b}$ = 15.7.\\
\begin{figure}[h!]
	
	\includegraphics[scale=0.32]{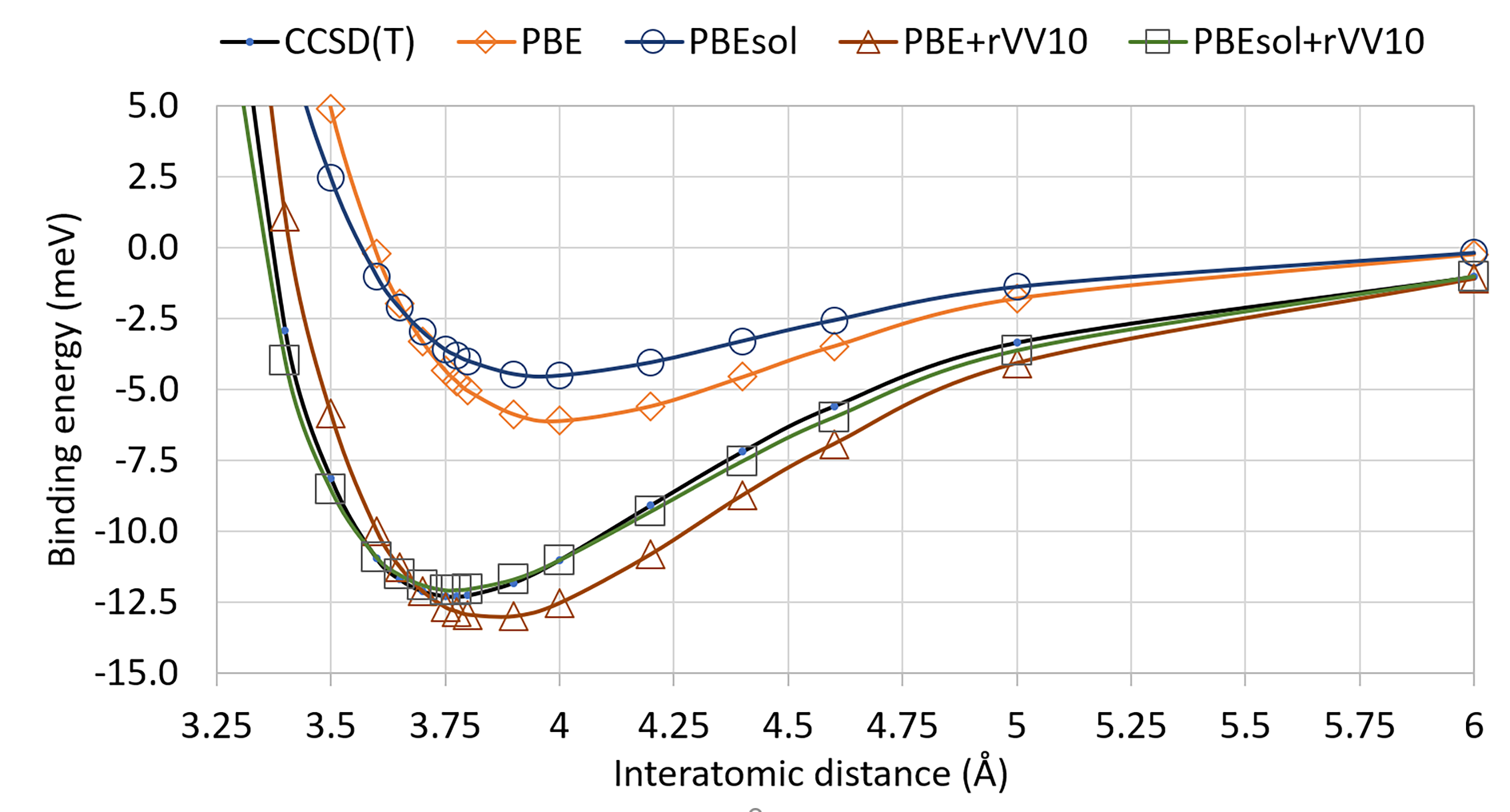} 
	\caption{The binding energy curve of the Ar dimer from PBE, PBEsol and their corresponding rVV10 corrected versions with respect to CCSD(T) curve. The value of $\textit{b}$ for PBE+rVV10 and PBEsol+rVV10  are 9.8 and 9.7, respectively.}
\end{figure}
For comparison we have combined the PBE$\cite{perdew1996generalized}$ and PBEsol$\cite{csonka2009assessing}$ GGA’s with the rVV10 correction shown in Figure 2. With MAPE of 1.76\%, the $\textit{b}$ parameter for PBEsol+rVV10 is found to be 9.7 while 9.8 is the $\textit{b}$ parameter determined for PBE+rVV10. Surprisingly PBEsol gives less binding than PBE. One of the reasons for the larger $\textit{b}$ parameter value for PBE+rVV10 is its inability to give the minimum position correctly. While all other methods such as SCAN+rVV10, revSCAN+rVV10 and PBEsol+rVV10 give the minimum around 3.775 {\AA} in agreement with CCSD(T), the PBE+rVV10 yields the minimum at around 4.0 {\AA}. 
Our results show that PBEsol+rVV10 gives the best fit to Argon dimer followed by revSCAN+rVV10 and SCAN+rVV10 while PBE+rVV10 gives the relatively worst fit.
\section{Computational details}
All the DFT calculations were performed using the projector-augmented wave (PAW) formalism implemented in the Vienna ab initio simulation package (VASP) code. The lattice constants of silver, gold and copper were obtained by the geometry relaxations of their respective bulk structures using different XC functionals. (4x4) supercell of  111, 110 and 100 surfaces using optimized lattice constants were built in the atomic simulation environment (ASE).$\cite{ase1,ase2}$ The supercell has a five-atomic-layer thickness. In order to prevent the interactions due to the periodic images, a vacuum of 12 {\AA} was added along the z-direction. To reduce the computational cost, the positions of the atoms on the bottom three layers were kept fixed, only allowing atoms on the two top layers to relax.
The thiophene molecule was constructed using the reference C-S, C-C and C-H bond lengths$\cite{harshbarger1970electron,mootz1981crystal}$ which was allowed to relax in the slab whose dimension was identical to that of the surface on which the adsorption occurs. Initially the thiophene was placed in a parallel orientation 3 {\AA} above the top metal layer and was allowed to relax. The center of mass and the azimuthal angle of the thiophene was used for the classification of the geometry. High symmetry sites, namely top, hollow, bridge, shortbridge, longbridge, fcc and hcp,$\cite{ase1,ase2}$ were used as sites for adsorption, e.g., top-45 indicates the center of mass of the thiophene adsorbed on top of the metal atom with a symmetry axis rotated by 45$^{\circ}$ from the direction of metal rows. The surface, the thiophene and the thiophene over the surface were all separately relaxed.
PAW potentials as recommended in the VASP manual were used for all the calculations. A plane wave cutoff of 650 eV and a thermal-smearing temperature of $k_{B}T$ = 0.1 eV were chosen for both the bulk and surface calculations. The Brillouin zone was sampled using a 4x4x1 Monkhorst-Pack k-point set for surfaces while 20x20x20 were used in the case of bulk relaxations. The structure of thiophene was optimized before adsorbing over one side of the slab (i.e., coverage of 1/16). Since both energies and equilibrium distances are dependent on the site of the adsorption, all major high symmetry sites were chosen for relaxing thiophene over metallic surfaces. The adsorption energy was calculated by subtracting the energy of the combined system (surface + thiophene) from the energy of the surface alone and the energy of the thiophene alone.
\begin{equation}
	E_{ads}= E_{surface+thiophene} - E_{surface} - E_{thiophene}
\end{equation}
\section{Results and discussion}
\subsection{Lattice constants of transition metals}
In the adsorption process, the organic molecule binds to the metal surface. A full picture about the exchange-correlation (XC) approximations utilized must include the lattice constants, which impact the geometry during the adsorption process.
We have assessed SCAN, revSCAN, PBE, PBEsol and their long-range van der Waals-corrected versions for the lattice constant of the three transition metals. Table 1 displays the lattice constants, which are compared to the zero-point-phonon corrected experimental lattice constants.$\cite{hao2012lattice}$\\
\begin{table}[h!]
	\centering
	\setlength{\tabcolsep}{20pt}
	\begin{tabular}{lccc}
		\hline
		& Copper & Silver & Gold \\
		\hline
		Reference$\cite{hao2012lattice}$ & 3.60  & 4.06  & 4.06 \\
		PBE   & 3.63  & 4.15  & 4.16 \\
		PBEsol & 3.57  & 4.05  & 4.08 \\
		SCAN  & 3.56  & 4.08  & 4.09 \\
		revSCAN & 3.57  & 4.11  & 4.11 \\
		PBE+rVV10 & 3.62  & 4.11  & 4.13 \\
		PBEsol+rVV10 & 3.55  & 4.02  & 4.06 \\
		SCAN + rVV10 & 3.54  & 4.06  & 4.07 \\
		revSCAN + rVV10 & 3.55  & 4.06  & 4.08 \\
		\hline
	\end{tabular}%
	\label{tab:addlabel}%
	\caption{The lattice constants (in {\AA}) of substrate metals, compared to the experimental zero-point phonon corrected lattice constants.}
\end{table}%
The results in Table 1 show that PBEsol systematically shortens the lattice constants by roughly about 0.07 {\AA} compared to the too-long PBE values making the PBEsol results quite accurate. Although SCAN is accurate for many bonding situations,$\cite{sun2016j}$ it is not particularly accurate for the lattice constants of the transition metals studied here. The revSCAN results are slightly longer than the SCAN values making the revSCAN values slightly better for copper but worse for silver and gold. The application of rVV10 shortens the lattice constants. For copper it improves the too-long PBE results but worsens the already shorter PBEsol, SCAN and revSCAN results. Though this shortening effect helps to obtain excellent latice constants for Gold and Silver with SCAN+rVV10, revSCAN+rVV10, it is not enough to correct the too-long PBE values. 
\subsection{Adsorption energies and geometry of thiophene on Cu(111), Ag(111) and Au(111)}
We have assessed the adsorption of the thiophene molecule on three crystal faces of copper, silver and gold, considering the adsorption energies, the adsorption geometry and the tilt angle between thiophene and the metal surface. Moving from Cu toward Au, the nature of the adsorption on these three metal surfaces changes. The adsorption on Cu(111) is a mixture of covalent and weak interactions, while the interaction on Au(111) is dominated by weak van der Waals interactions $\cite{imanishi1998structural,milligan2001complete,elfeninat1995theoretical,liu2002chemistry}$. Though, the experiments do not give the precise value of adsorption energies$\cite{milligan2001complete}$, they $\cite{sexton1985vibrational,liu2002chemistry}$ overall report a strong dependence on the coverage of the thiophene adsorption. The structural information of the adsorbed molecule on the metallic surface such as molecule-surface distance, the angle of the adsorbed molecule and surface, and the adsorption sites vary with increasing coverage.\\
Irrespective of the exchange-correlation functional and vdW correction applied, the adsorption of thiophene on (111) surfaces of the different metals displays some common features. Based on the given coverage, our calculations find that the fcc-45 is the most stable site of adsorption for all metals. The same adsorption site was found to be the most stable with our benchmark PBE+vdW-dZK approximation, and experiments support these results too. The predicted fcc-45 adsorption site for Cu(111) is close to the top adsorption site predicted by experiments for Cu(111)$\cite{milligan2001complete,rousseau2002structure}$. The difference is just the result of the choice of  the position of the reference point in thiophene.\\
According to the experimental results, an increased tilt angle is observed with increasing coverage, while lower coverage prefers  a slightly lower tilt angle$\cite{milligan2001complete}$ for thiophene adsorbed on Cu(111). Though a higher tilt angle of 55$^{\circ}$ was found at a significantly higher coverage of thiophene on Au(111) than ours,$\cite{nambu2003film,dishner1996formation}$ experiments indicate the preference of thiophene lying flat on both Ag(111) and Au(111) at low coverage.$\cite{liu2002chemistry,vaterlein2000orientation,chen1996spatially}$\\
Our PBE+vdW-dZK method gives a tilt angle of 7$^{\circ}$ - 17$^{\circ}$ for thiophene adsorbed on Cu(111) surface, a tilt angle of 1$^{\circ}$ - 2$^{\circ}$ for thiophene adsorbed on Ag(111) and almost zero tilt angle on Au(111). Our other methods give a tilt angle of 1$^{\circ}$ - 6$^{\circ}$ for thiophene adsorbed on Cu(111), a tilt angle of 1$^{\circ}$ - 2$^{\circ}$ on Ag(111), and tilt angles of 1$^{\circ}$ - 4$^{\circ}$ on Au(111). \\Experiments$\cite{milligan2001complete}$ for adsorption of thiophene over copper suggest a range of tilt angle of 20$^{\circ}$$\pm$3$^{\circ}$ at the coverage of 0.05 ML, and 25$^{\circ}$$\pm$4$^{\circ}$ at the coverage of 0.1 ML. Our PBE+vdW-dZK results giving the tilt angle of 17$^{\circ}$ at the most stable site at the coverage of 0.0625 ML, agree very well with the experiments.\\
The minimum Cu-S distance of 2.57 {\AA} that we show in Table 2 from the PBE+vdW-dZK method based on LZK theory$\cite{tao2018modeling}$ is in close agreement to the experiments.$\cite{milligan2001complete}$ Though both SCAN and revSCAN predict slightly longer Cu-S distance, SCAN+rVV10 and revSCAN+rVV10 results are closer to the experiments.\\
We have not found precise experimental data for the distance between the adsorbed S atom of thiophene and the Ag(111) and Au(111) surface. Both SCAN and revSCAN and the corresponding rVV10 corrected methods yield almost identical adsorption distance irrespective of whether the surface is Ag(111) or Au(111). The PBE+vdW-dZK method gives a slightly larger distance of 3.23 {\AA} for Au(111) and 3.16 {\AA} for Ag(111), compared to other methods discussed here. However, the latter result gives a very good match with the previously studied$\cite{maurer2016adsorption}$ PBE-vdW$^{surf}$ method for the distance of Ag and S atoms for Ag(111). The adsorption distance predicted by SCAN+rVV10 is close to the results of Maurer et al.$\cite{maurer2016adsorption}$ for thiophene adsorbed on Au(111).\\
\begin{figure}[h!]
	
	\includegraphics[scale=0.33]{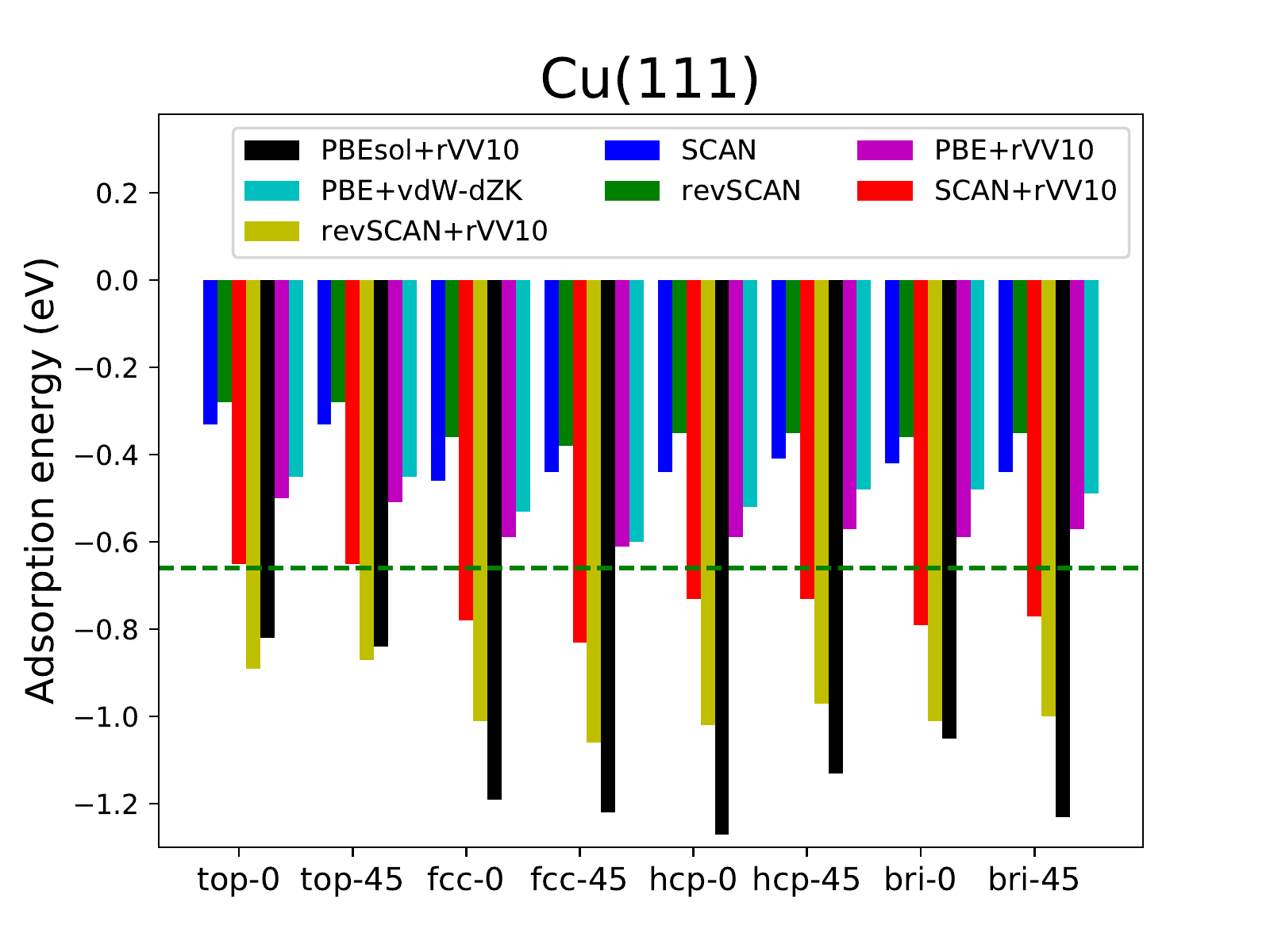}
	\includegraphics[scale=0.33]{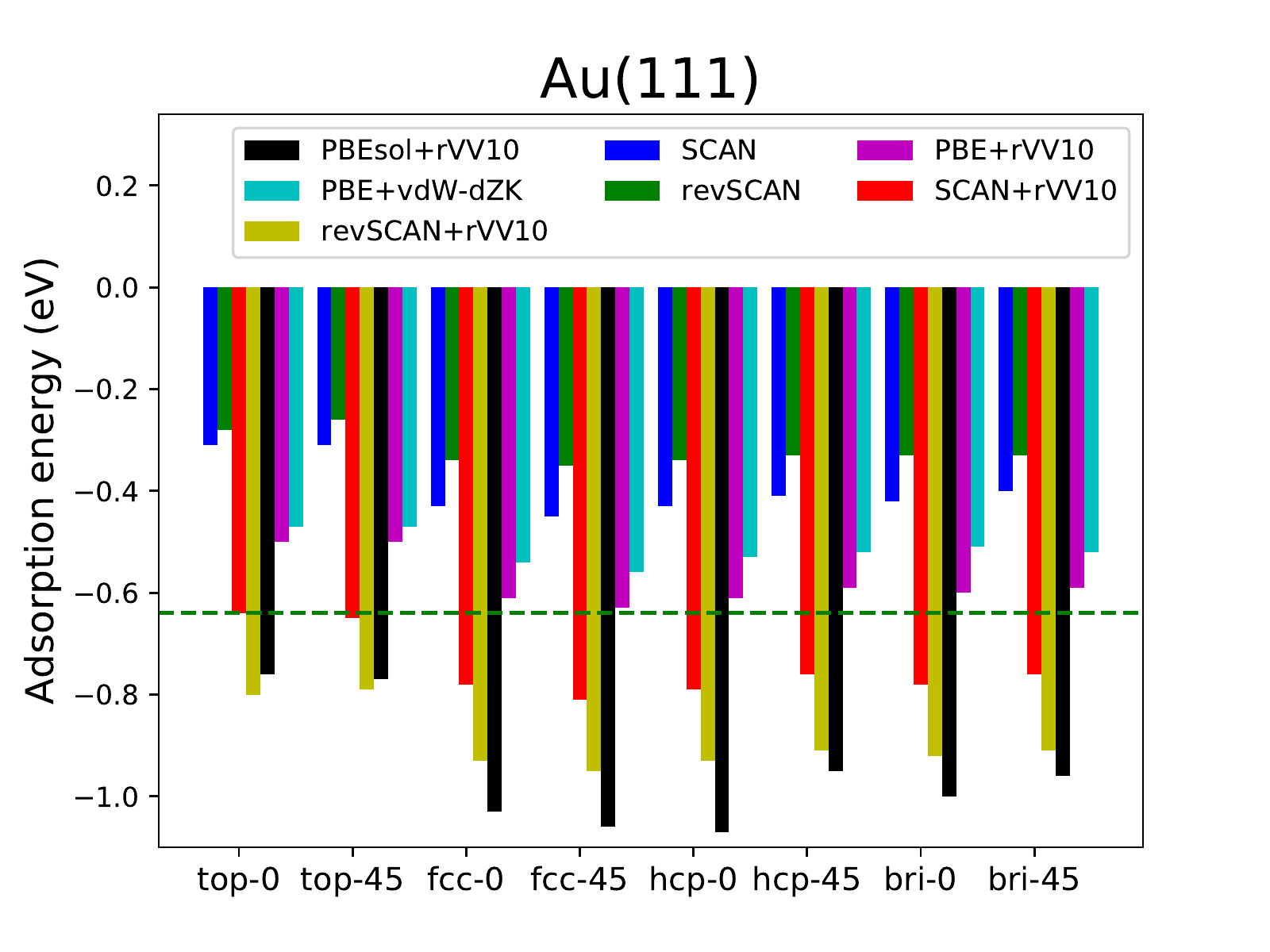} 
	\includegraphics[scale=0.33]{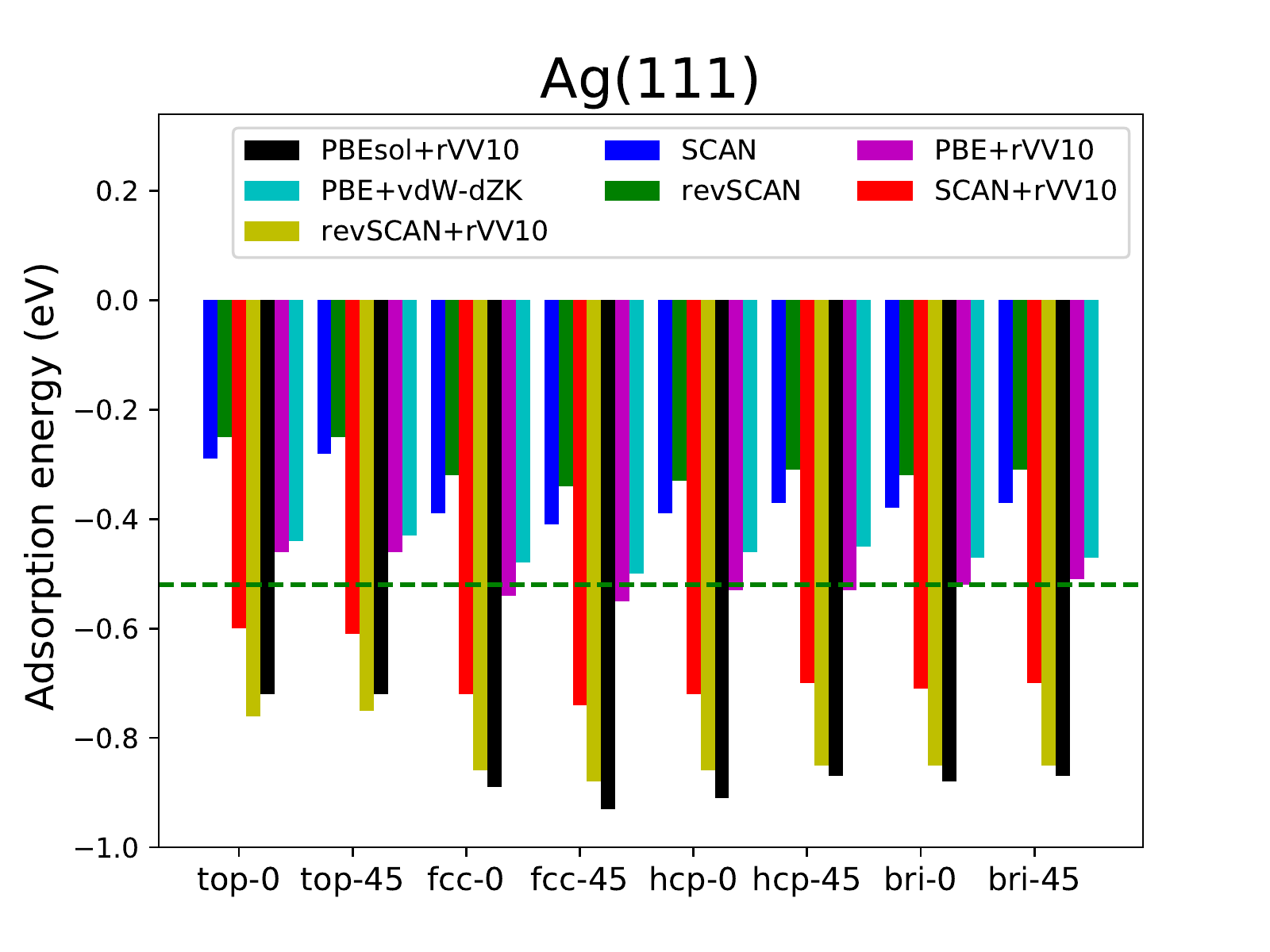}
	\caption{The SCAN, revSCAN, SCAN+rVV10, revSCAN+rVV10, PBE+rVV10, PBEsol+rVV10 and PBE+vdW-dZK adsorption energies with respect to two hollow sites namely fcc and hcp, bridge site denoted by bri and top site (the numbers show rotation angle values) over Cu(111), Au(111) and Ag(111). The dashed green horizontal line is the reference experimental adsorption energy.$\cite{milligan2001complete,vaterlein2000orientation,liu2002chemistry}$ }
	
\end{figure}
\begin{table}[h!]
	\centering
	\setlength{\tabcolsep}{10pt}
	\begin{tabular}{ccccccc}
		\hline
		\multirow{2}[0]{*}{} & \multicolumn{2}{c}{Cu} & \multicolumn{2}{c}{Ag} & \multicolumn{2}{c}{Au} \\
		\hline
		& d(Cu-S) & d(C-S) & d(Cu-S) & d(C-S) & d(Cu-S) & d(C-S) \\
		\hline
		Expt$\cite{harshbarger1970electron,mootz1981crystal,milligan2001complete}$  & 2.62$\pm$0.03    & 1.71  &   -   & 1.71  & -     & 1.71 \\
		PBE+vdW-dZK & 2.57  & 1.71  & 3.16  & 1.71  & 3.23  & 1.71 \\
		SCAN  & 2.70  & 1.71  & 3.00  & 1.71  & 2.98  & 1.71 \\
		revSCAN & 2.88  & 1.71  & 3.03  & 1.70  & 3.02  & 1.70 \\
		SCAN + rVV10 & 2.57  & 1.71  & 2.97  & 1.71  & 2.93  & 1.71 \\
		revSCAN + rVV10 & 2.56  & 1.71  & 2.93  & 1.71  & 2.91  & 1.71 \\
		PBE + rVV10 & 2.88  & 1.71  & 3.00  & 1.71  & 2.98  & 1.71 \\
		PBEsol + rVV10 & 2.19  & 1.71  & 2.68  & 1.71  & 2.59  & 1.71 \\
		\hline
	\end{tabular}%
	\label{tab:addlabel}%
	\caption{Distances (in {\AA}) between the sulphur atom in thiophene and the nucleus of the nearest atom of the metal surface. The PBE+vdW-dZK method was presented in ref. 40 and 45.}
\end{table}%
\begin{table}[h!]
	\centering
	\setlength{\tabcolsep}{30pt}
	\begin{tabular}{cccl}
		\multirow{2}[0]{*}{} & \multicolumn{3}{c}{} \\
		\hline
		& Cu    & Ag    & Au \\
		\hline
		Expt$\cite{milligan2001complete,liu2002chemistry,vaterlein2000orientation}$  & -0.66 & -0.52 & -0.64 \\
		PBE+vdW-dZK & -0.60 & -0.50 & -0.56 \\
		SCAN  & -0.44 & -0.41 & -0.45 \\
		revSCAN & -0.38 & -0.34 & -0.35 \\
		SCAN + rVV10 & -0.83 & -0.74 & -0.81 \\
		revSCAN + rVV10 & -1.06 & -0.88 & -0.95 \\
		PBE + rVV10 & -0.61 & -0.55 & -0.63 \\
		PBEsol + rVV10 & -1.22 & -0.93 & -1.06 \\
		\hline
	\end{tabular}%
	\label{tab:addlabel}%
	\caption{Adsorption energy (in eV) of thiophene on Cu(111), Ag(111) and Au(111) surfaces with GGA and meta-GGA-based approximations with respect to the most stable fcc-45 adsorption site. The estimated (see the supplementary information for more details) adsorption energies ($\pm$ 0.2 eV)$\cite{christian2016surface,liu2015quantitative}$ from the TPD temperature maxima (Expt.),$\cite{milligan2001complete,liu2002chemistry,vaterlein2000orientation}$ and the results from the PBE+vdW-dZK model$\cite{tao2018modeling}$ are also shown.}
\end{table}%
The results in Table 3 show that our theoretical benchmark the PBE+vdW-dZK approximation adsorption energies are in a good agreement with the experimental  adsorption energies estimated from TPD temperature maxima $\cite{milligan2001complete,liu2002chemistry,vaterlein2000orientation}$  using the Redhead's model$\cite{de1990thermal}$ (see the supplementary information for more details). Our analysis shows that the adsorption energies estimated properly from Redhead's model are considerably more precise for thiophene and benzene adsorption on coinage metal surfaces than suggested before$\cite{christian2016surface,liu2015quantitative}$ (cf. supplementary information). Notice the slightly different coverage given in most of the experiments.$\cite{milligan2001complete,imanishi1998structural}$ Both SCAN and revSCAN underbind thiophene on Cu(111), Ag(111), and Au(111) surfaces compared to our theoretical reference.  However, with the added rVV10 corrections they overbind. SCAN+rVV10  works less well for the thiophene adsorption on Cu(111) than for the adsorption of benzene.$\cite{peng2016versatile}$ revSCAN by construction was designed to remove some of the intermediate range interactions of SCAN, so an underbinding of thiophene on coinage metal surfaces is expected, but surprisingly revSCAN+rVV10 is more overbinding than SCAN+rVV10. The reason behind the strong overbinding of revSCAN+rVV10 is the inclusion of relatively larger vdW correction through smaller $\textit{b}$ value. SCAN and in particular revSCAN are significantly underbinding for adsorption of thiophene on Ag(111) and Au(111) surfaces too, but adding the rVV10 corrections again leads to overbinding. The SCAN+rVV10 is overbinding  compared to the earlier results from  PBE+vdW$^{surf}$ $\cite{maurer2016adsorption}$ and B86bPBE-XDM approximations$\cite{christian2016surface}$ too. Comparison with the relevant results of Christian et al.$\cite{christian2016surface}$ shows that B86bPBE-XDM results do not reflect the qualitative tendency that Cu and Au surfaces bind the thiophene about equally strongly and slightly stronger than Ag. This tendency is reproduced by all methods in Table 3, and especially well quantitatively reproduced by PBE+vdW-dZK, SCAN+rVV10, and PBE+rVV10.\\
For comparative purposes we show PBE+rVV10 and PBEsol+rVV10 results in Table 3. PBEsol is known to contain a certain amount of medium-range correlation compared to PBE. The combination of these approximations with rVV10 can serve as a simplified model of the more sophisticated meta-GGA’s with rVV10. Due to the lower-order gradient correction in the correlation energy and the decreased medium-range enhancement in its exchange the revSCAN meta-GGA resembles PBE, while SCAN exhibits more analogy with PBEsol. Being inspired by this analogy, we have computed the adsorption energy, distances and tilting of thiophene with PBE+rVV10 and PBEsol+rVV10.\\
Surprisingly PBE+rVV10 is more reliable than SCAN+rVV10 or revSCAN+rVV10. The adsorption energy on Cu(111) is overestimated by revSCAN+rVV10 and SCAN+rVV10, while PBE+rVV10 predicts less overbinding. This reliability of PBE+rVV10 is present for Ag(111) and Au(111). Adsorption energies on Ag(111) and Au(111) from PBE+rVV10 not only agree with the PBE+vdW-dZK results, but are very close to the estimated experimental values. The PBEsol+rVV10 approximation, although it is remarkably accurate for diverse properties including the binding energy of Xe on Cu(111) and Ag(111),$\cite{terentjev2018dispersion}$ turns out less successful in the case of adsorption of thiophene on Cu(111), Ag(111) and Au(111). PBEsol+rVV10 predicts too large binding energies and too short adsorption distances. The PBE+rVV10 method however is unable to yield the moderate tilting of thiophene over Cu(111), and is able to predict the almost parallel orientation over Ag(111) and Au(111) surfaces. The predicted adsorption distance from PBE+rVV10 is slightly longer than the value predicted by the experiments$\cite{milligan2001complete}$ for thiophene over Cu(111). We are in lack of an accurate value of adsorption distance from experiments for thiophene over Ag(111) and Au(111) surfaces, but PBE+rVV10 values  agree with earlier results from the $PBE+vdW^{Surf}$ method.$\cite{maurer2016adsorption}$\\
It is more surprising that revSCAN+rVV10 significantly overbinds thiophene on all three transition metals, more than SCAN+rVV10 does. The na\"ive expectation from the combination of revSCAN and rVV10 is a more balanced description of weak interaction. The overall conclusion is that the combination of SCAN, revSCAN and rVV10 does not work accurately in general. A similar conclusion was drawn about the SCAN+MBD approximation.$\cite{hermann2018electronic}$ When the MBD method was combined with SCAN, the effective range of SCAN depended on system size.\\
Such long-range corrections need a different empirical cutoff parameter for each semi-local functional, in order to avoid a misrepresentation of intermediate-range vdW interaction. The pairing of semi-local and nonlocal terms can work well for some systems and fail for others.$\cite{hermann2018electronic}$ In particular, the pairing of SCAN with rVV10 works well for layered materials and for a benzene molecule adsorbed on coinage metals. But for liquid water, for some molecular crystals, and for the problem considered here, adsorption of thiophene on transition metals, this pairing overbinds. A familiar proposed solution would be to start from a semi-local functional that has little (PBE) or no (revSCAN) intermediate-range vdW interaction, and get the intermediate-range contribution from rVV10. This solution is consistent with the fact that semi-local functionals yield intermediate-range vdW attraction from their exchange energy terms. But we show here that this solution does not always work.
While the PBE based vdW corrected methods performed better, revSCAN+rVV10 performed poorly. Both PBE+vdW-dZK and PBE+rVV10 methods gave overall better adsorption energies. PBE+vdW-dZK in particular was outstanding. It not only predicted the adsorption energies reasonably well, but also predicted the adsorption distance and tilting of the thiophene correctly.\\ 
Our results for the Cu, Ag and Au (100) and (110) surfaces can be found in the Supplementary Material. In general, the adsorption of thiophene on the (100) and (110) crystal faces is less explored experimentally. With all methods applied in our work, except SCAN we generally observe a stronger overbinding of the thiophene molecule on the Cu(100) than on Cu(111) surface.$\cite{orita2004adsorption,imanishi1996structural,sony2007importance,malone2017van}$ Considering the hollow-45 site as the most stable adsorption site, the SCAN and revSCAN adsorption energy of -0.89 eV and -0.76 eV are reasonably accurate even without vdW correction, while SCAN+rVV10 and revSCAN+rVV10 both overbind. The revPBE-vdW and optB88-vdW approximation were found give the similar trend predicting -0.50 eV and -0.85 eV binding energy, respectively.$\cite{sony2007importance,malone2017van}$ Among the approximations that we have applied, the revSCAN approaches these values the best, giving -0.76 eV binding energy at the hollow-45 site.
\subsection{Surface energies of the three transition metals with GGA and meta-GGAs, and their vdW-corrected versions}
Practically, due to cancellation in Eq. 7, the surface energy makes a minor contribution to the adsorption energy. Independent of the adsorption energies, the surface energies obtained by GGA and meta-GGA-based methods and their vdW corrections can still shed some more light on how the long-range vdW correlation works with the short-range exchange and correlation. Recently, Patra et al.$\cite{patra2017properties}$ demonstrated that the long-range vdW correction has a significant role in the surface energy. The surface energy is evaluated from the energies of the bulk and the slab as:
\begin{equation}
	\sigma = \lim_{n\to\infty} \frac{E(n)-nN\epsilon_{bulk}}{2A}\\
\end{equation}
with the information of the surface area A and n as the number of layers. \\
\begin{table}[h!]
	\centering
	\resizebox{\columnwidth}{!}{%
		\begin{tabular}{lccccccccr}
			\hline
			Metal & PBE   & PBEsol & SCAN  & revSCAN & PBE+rVV10 & PBEsol+rVV10 & SCAN+rVV10 & revSCAN+rVV10 & \multicolumn{1}{c}{Reference$\cite{tyson1975,tyson1977,patra2017properties}$} \\
			\hline
			Cu    & 1.41  & 1.76  & 1.71  & 1.73  & 1.74  & 2.11  & 1.92  & 2.07  & \multicolumn{1}{c}{1.79 $\pm$ 0.19} \\
			Au    & 0.82  & 1.13  & 1.06  & 1.06  & 1.17  & 1.52  & 1.29  & 1.43  & \multicolumn{1}{c}{1.51 $\pm$ 0.16} \\
			Ag    & 0.80  & 1.09  & 1.04  & 1.04  & 1.11  & 1.43  & 1.24  & 1.36  & \multicolumn{1}{c}{1.25 $\pm$ 0.13} \\
			\\
			ME    & -0.51 & -0.19 & -0.25 & -0.24 & -0.18 & 0.17  & -0.03 & 0.10  &  \\
			MAE   & 0.51  & 0.19  & 0.25  & 0.24  & 0.18  & 0.17  & 0.12  & 0.15  &  \\
			\hline
		\end{tabular}}
		\label{tab:addlabel}%
		\caption{Computed surface energies (in J/m$^2$) of different transition metals}
	\end{table}%
	Our PBE, PBEsol, SCAN and SCAN+rVV10 results displayed in Table 4 agree with the surface energies found in ref.$\cite{tyson1975,tyson1977,patra2017properties}$ The results in Table 4 show that PBE seriously underestimates the surface energies in agreement with the results for jellium surface energies.$\cite{csonka2009assessing}$ This underestimation is well corrected by the rVV10 terms for Cu but not for Ag and Au. PBEsol, SCAN and revSCAN all yield excellent results for Cu, but not for Ag and Au. Addition of rVV10 corrections to PBEsol, SCAN and revSCAN worsen considerably the already excellent surface energies for Cu but yield better surface energies for Ag and Au. The best overall performer is SCAN+rVV10 followed by revSCAN+rVV10. These results suggest that, the Cu surface energy can be accurately calculated with a proper GGA or metaGGA without any rVV10 corrections, but accurate surface energies for Ag and Au require rVV10 corrections.	
	\section{Conclusion}
	To describe the equilibrium adsorption of thiophene on coinage metal surfaces, we selected several promising density functionals such as SCAN, revSCAN, PBE and PBEsol.  As GGA and meta-GGA functionals miss the long-range dispersion, we added rVV10  corrections as already suggested for SCAN. It is known that  pairwise-interaction models such as rVV10 show incorrect asymptotic behavior, but as it was shown earlier this error is negligible around the equilibrium distance.\\  
	rVV10 correction requires one parameter $\textit{b}$ to adjust the long range part of interaction to the range of  interaction present in the given parent functional. To obtain a fitted value we have chosen for reference the calculated CCSD(T) potential energy curve for Ar dimer. This curve is in reasonable agreement with the experiment and with higher level CCSDT(Q) curves around the well and the attractive branch of the curve.  Our results for combining rVV10 to revSCAN, PBE, and PBEsol are $\textit{b}$ = 9.4, 9.8, and 9.7, respectively. With these parameters the revSCAN+rVV10 and PBEsol+rVV10 reproduce correctly the minimum energy around the correct 3.77 {\AA} equilibrium distance, however the PBE+rVV10 yields minimum around 4 $^{\circ}$. The order from best to worst fit is the following: PBEsol+rVV10, revSCAN+rVV10, SCAN+rVV10, and PBE+rVV10. In our calculations we apply the original b = 15.7 for SCAN+rVV10.\\
	Our results for the lattice constants for Cu, Ag, and Au show that SCAN is not as accurate as PBEsol that yields the best lattice constants.  The SCAN and revSCAN lattice constants agree well with each other, too short for Cu and too-long for Ag and Au. After the rVV10 correction the lattice constants shorten by 0.1-0.2 {\AA} for Cu and by 0.2-0.5 {\AA} for Ag and Au. The PBE+rVV10 lattice constants remain too-long and the PBEsol+rVV10 lattice constants become too short except for Au. The rVV10 corrected SCAN and revSCAN lattice constants are too short for Cu, excellent for Ag, and 0.1-0.2 {\AA} too-long for Au.\\
	We have shown that PBE seriously underestimates the  surface energies of Cu, Ag, and Au, in agreement with the results obtained for jellium surface energies. This underestimation can be remedied by rVV10 correction for Cu but not for Ag and Au. PBEsol, SCAN and revSCAN  yield excellent surface energies for Cu, but rVV10 corrections worsen the results.\\ 
	Our results show that the lowest energy adsorption site on the coinage metal (111) surfaces is fcc-45 by all methods used in this paper, in agreement with experiment.  For metal thiophene distances and adsorption energies we have chosen the PBE+vdW-dZK  method for reference as it mimics the very accurate but computationally too demanding RPA binding energies perfectly for the interaction of graphene and metal surfaces. For the Cu-S distance revSCAN+rVV10, SCAN+rVV10 yield good agreement with our reference method and with the experiment.\\
	According to our calculations SCAN and revSCAN underbind thiophene on Cu(111), Ag(111), and Au(111) surfaces by 0.1-0.3 eV. The rVV10 correction adds 0.3-0.5 eV to the binding energy making revSCAN+rVV10, SCAN+rVV10 overbinding by 0.2-0.4 eV. PBE+vdW-dZK and PBE+rVV10 show excellent agreement with estimated experimental results. PBEsol+rVV10 yields serious (0.4-0.6 eV) overbinding in accordance with the too short metal-S distance. Our calculations reflect the qualitative tendency that Cu and Au surfaces bind the thiophene about equally strongly and slightly stronger than Ag surface. This tendency is quantitatively reproduced by PBE+vdW-dZK, SCAN+rVV10, and PBE+rVV10.\\
	We have demonstrated that good results of the rVV10 corrected density functionals for the well depth and the attractive region of the Ar dimer dissociation curve do not guarantee good results for thiophene adsorption on coinage metals. The order from best to worst fit is the following: PBEsol+rVV10, revSCAN+rVV10, SCAN+rVV10, and PBE+rVV10. However, the order of the performance for thiophene adsorption is the opposite with PBE+rVV10 being the best and PBEsol+rVV10 being the worst. We clearly show that the good fit to the Ar dimer curve does not guarantee good adsorption energies of polar molecules, e.g., thiophene on coinage metals.\\
	The best method for thiophene adsorption is PBE+vdW-dZK, which is not only quantitatively correct for the adsorption energies but also predicts the ordering of adsorption energies for copper, gold and silver right along with the tilting angles and adsorption distances in good agreement with the experiment.
	\begin{acknowledgement}
		We acknowledge the donors of the American Chemical Society Petroleum Research Fund for support of this research. We thank Center for Computational Design of Functional Layered Materials (CCDM) and High Performance Computing (HPC) at Temple University for providing the computational facilities. We thank John P. Perdew, Haowei Peng, Jefferson E. Bates, Hemanadhan Myneni, and Niraj K. Nepal for their valuable help during the calculations.
	\end{acknowledgement}

\clearpage
\section{Supporting info:}
	\begin{table}[h!]
		\centering
		\renewcommand\thetable{S1}
		\begin{tabular}{cccl}
			\hline
			& Cu    & Ag    & Au \\
			\hline
			Reference$\cite{sexton1985vibrational}$ & -0.63 & -     & - \\
			SCAN$\cite{sun2015strongly}$  & -0.89 & -0.49 & -0.64 \\
			revSCAN$\cite{mezei2018simple}$ & -0.76 & -0.40 & -0.45 \\
			SCAN + rVV10$\cite{sabatini2013nonlocal}$ & -1.30 & -0.82 & -1.01 \\
			revSCAN + rVV10 & -1.47 & -0.94 & -1.07 \\
			\hline
		\end{tabular}%
		\label{}
		\caption{Adsorption energy (in eV) of thiophene over Cu(100), Ag(100), and Au(100) surfaces with the meta-GGA based approximations and their corresponding vdW corrected versions with respect to the most stable adsorption site. Hollow-45 (the number 45 represents the value of angle of rotation) is found to be the most stable site of adsorption in 100 surfaces irrespective of the nature of metals and exchange-correlation(XC) functionals used.   }
	\end{table}%
	
	\begin{table}[h!]
		\centering
		\renewcommand\thetable{S2}
		\begin{tabular}{ccccccc}
			\hline
			\multirow{2}[0]{*}{} & \multicolumn{2}{c}{Cu} & \multicolumn{2}{c}{Ag} & \multicolumn{2}{c}{Au} \\
			\hline
			& d(Cu-S) & d(C-S) & d(Cu-S) & d(C-S) & d(Cu-S) & d(C-S) \\
			\hline
			Reference$\cite{harshbarger1970electron,imanishi1996structural}$ & -     & 1.71  & -     & 1.71  & -     & 1.71 \\
			SCAN  & 2.13  & 1.77  & 2.89  & 1.71  & 2.54  & 1.72 \\
			revSCAN & 2.17  & 1.76  & 2.97  & 1.70  & 2.98  & 1.70 \\
			SCAN + rVV10 & 2.13  & 1.77  & 2.75  & 1.71  & 2.45  & 1.72 \\
			revSCAN + rVV10 & 2.14  & 1.77  & 2.80  & 1.71  & 2.72  & 1.71 \\
			\hline
		\end{tabular}%
		\label{tab:addlabel}%
		\caption{Adsorption distances (in {\AA}) between the sulfur atom in thiophene and the nucleus of the nearest atom of the metal surface (100 surface) and the C-S bond length (in {\AA}). Measurements are made on the hollow-45 site, which is the most stable site of adsorption in 100 metal surfaces.}
	\end{table}%

	\begin{table}[h!]
		\centering
		\renewcommand\thetable{S3}
		\begin{tabular}{cccl}
			\hline
			& Cu    & Ag    & Au \\
			\hline
			SCAN  & -0.74 & -0.49 & -0.74 \\
			revSCAN & -0.53 & -0.38 & -0.49 \\
			SCAN + rVV10 & -1.12 & -0.77 & -1.05 \\
			revSCAN + rVV10 & -1.23 & -0.87 & -1.06 \\
			\hline
		\end{tabular}%
		\label{tab:addlabel}%
		\caption{Adsorption energy (in eV) of thiophene on Cu(110), Ag(110), and Au(110) surfaces with meta-GGA based approximations and their corresponding vdW corrected versions, with respect to the most stable site of adsorption. Shortbridge-45 is found to be the most stable site of adsorption irrespective of the nature of metal, except for revSCAN which predicts longbridge-0 as the most stable site of adsorption. We did't found the results from experiments to compare to our results.}
	\end{table}%
	
	\begin{table}[h!]
		\centering
		\renewcommand\thetable{S4}
		\begin{tabular}{ccccccc}
			\hline
			\multirow{2}[0]{*}{} & \multicolumn{2}{c}{Cu} & \multicolumn{2}{c}{Ag} & \multicolumn{2}{c}{Au} \\
			\hline
			& d(Cu-S) & d(C-S) & d(Cu-S) & d(C-S) & d(Cu-S) & d(C-S) \\
			\hline
			Reference$\cite{harshbarger1970electron}$ & -     & 1.71  & -     & 1.71  & -     & 1.71 \\
			SCAN  & 2.24  & 1.71  & 2.36  & 1.71  & 2.73  & 1.71 \\
			revSCAN & 2.43  & 1.71  & 2.53  & 1.7   & 2.82  & 1.71 \\
			SCAN + rVV10 & 2.24  & 1.71  & 2.35  & 1.71  & 2.67  & 1.72 \\
			revSCAN + rVV10 & 2.27  & 1.72  & 2.48  & 1.71  & 2.66  & 1.71 \\
			\hline
		\end{tabular}%
		\label{tab:addlabel}%
		\caption{Adsorption distances (in {\AA}) between the atom in thiophene and the nucleus of the nearest atom of the metal surface (110 surface) and the C-S bond length (in {\AA}). Measurements are made on the most stable site of adsorption in 110 metal surfaces. }
	\end{table}%
	\begin{figure}[h!]
		\renewcommand\thefigure{S1}
		\includegraphics[scale=0.5]{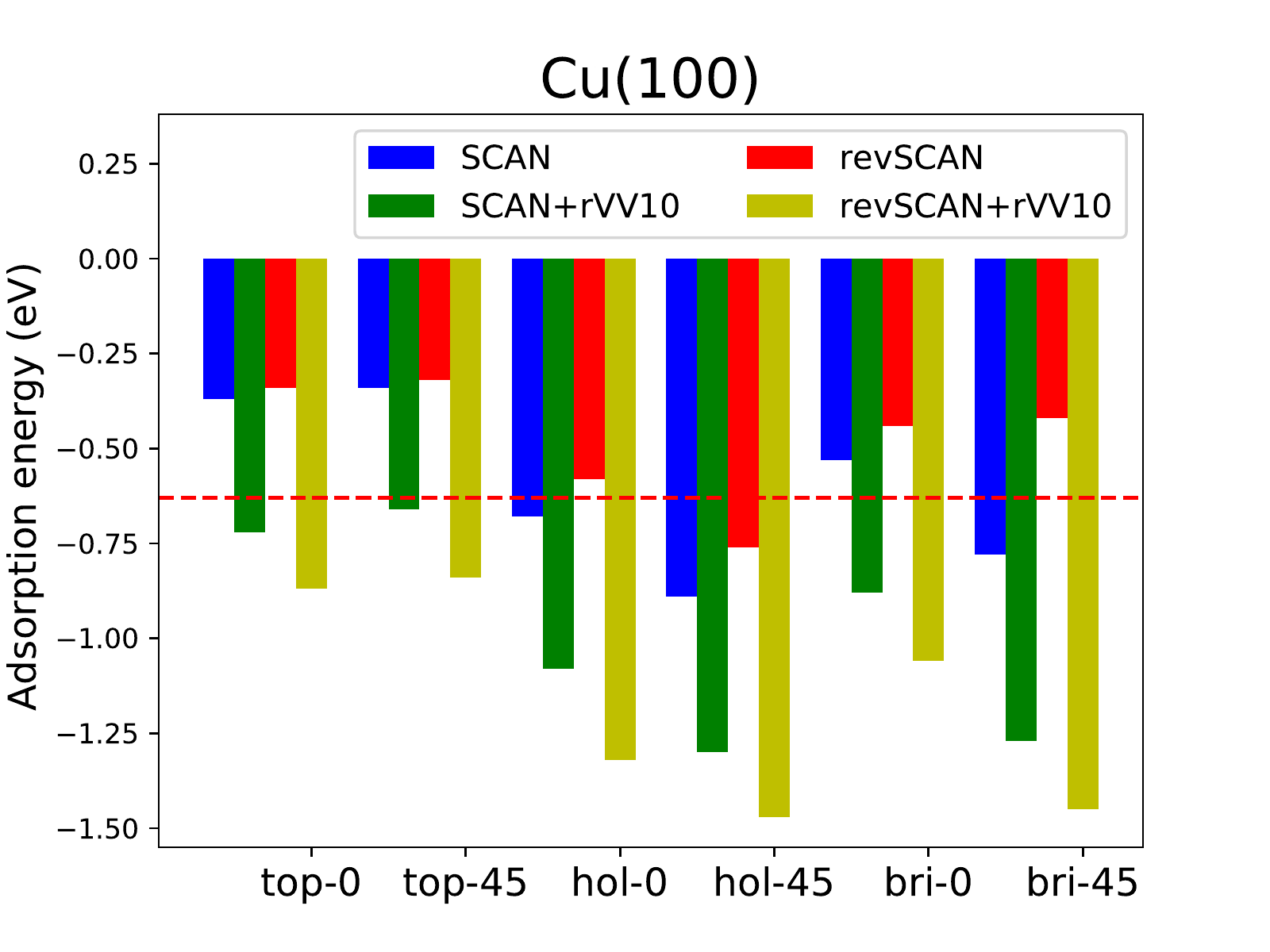} 
		\includegraphics[scale=0.5]{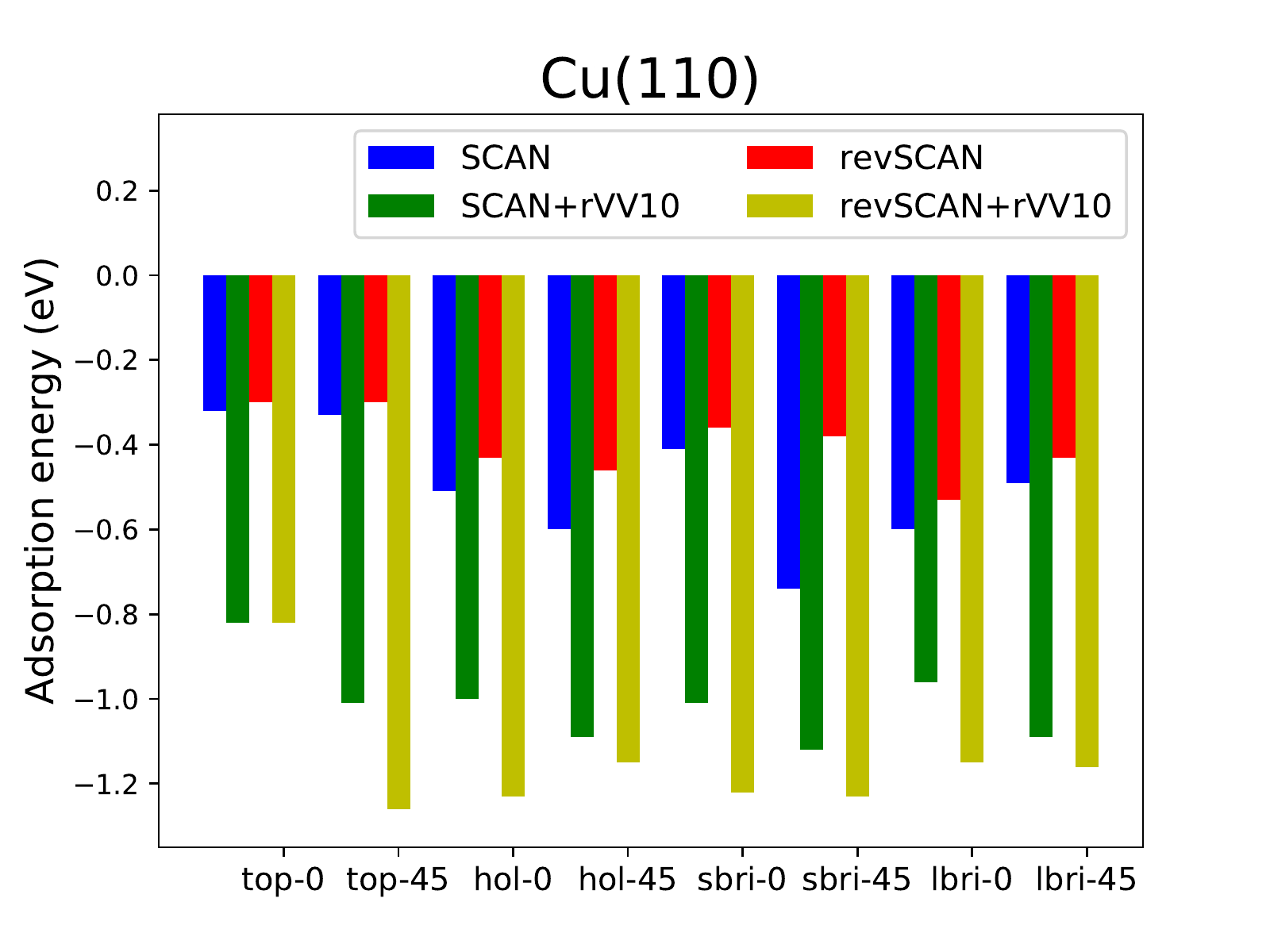} 
		\caption{(on left) Compares the site dependence of thiophene over Cu(100) surface. Hollow-45 site denoted by hol-45 is the most stable site of adsorption. The dashed horizontal red line shows the reference binding energy$\cite{imanishi1996structural}$. (on right) Compares the site dependence of thiophene on Cu(110) surface. All the XC functionals except revSCAN predicts shortbridge-45 (denoted by sbri-45) as the most stable site of adsorption. revSCAN predicts longbridge-0 (denoted as lbri-0) as the most stable adsorption site. The extent to which adsorption energy depeneds in site is stronger in copper compared to silver and gold.  }
	\end{figure}
	\begin{figure}[h!]
		\renewcommand\thefigure{S2}
		\includegraphics[scale=0.5]{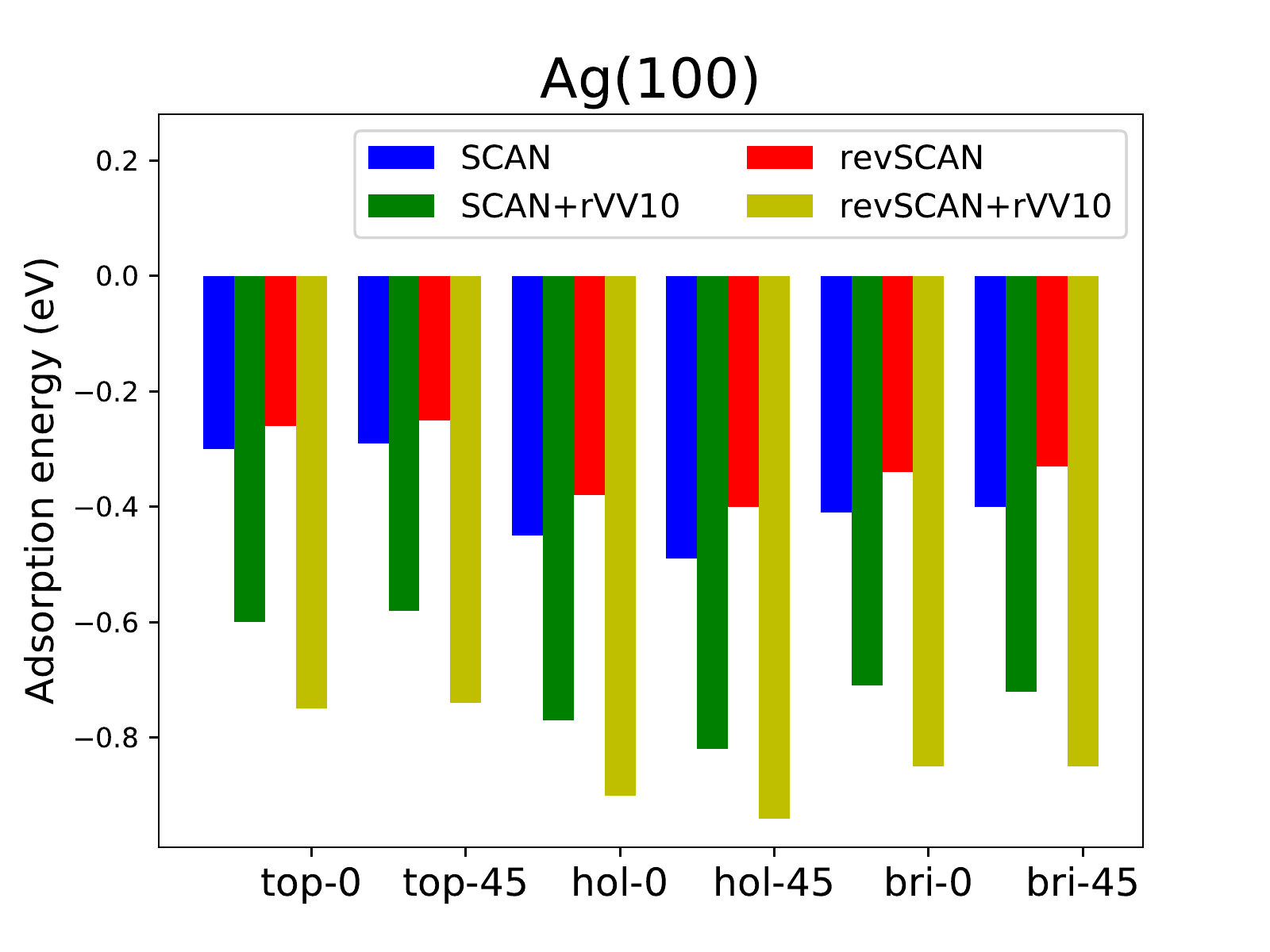} 
		\includegraphics[scale=0.5]{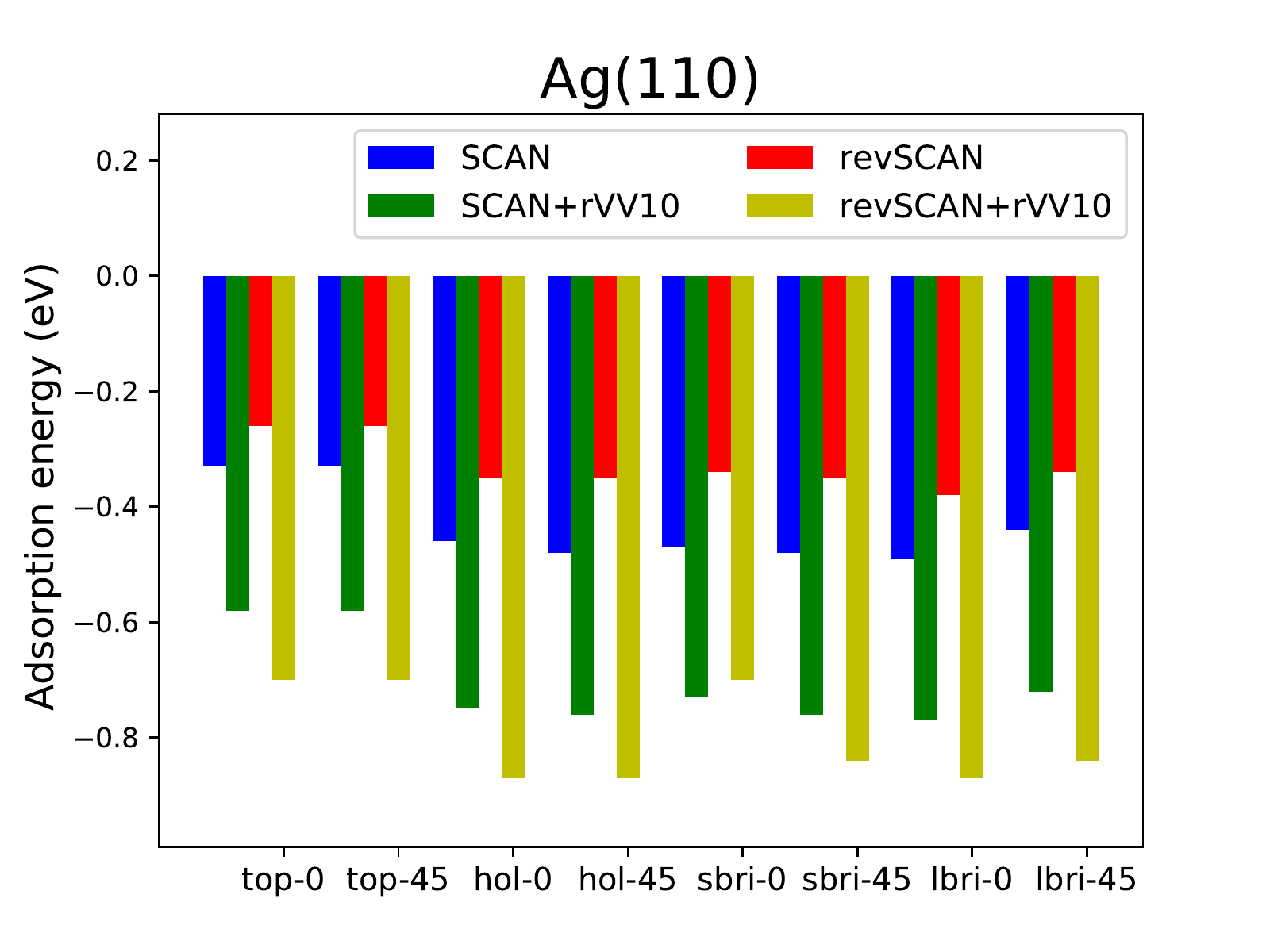} 
		\caption{(on left) Compares the site dependence of thiophene over Ag(100) surface. Hollow-45 site is the most preferred site of adsorption. (on right) Compares the site dependence of thiophene on Ag(110) surface. All the XC functionals except revSCAN predicts shortbridge-45 as the most stable site of adsorption. revSCAN predicts longbridge-0 as the most stable site. The extent of site dependence in silver is similar to that of gold but not as noticeable as it is in the case of copper. }
	\end{figure}	
	\begin{figure}[h!]
		\renewcommand\thefigure{S3}
		\includegraphics[scale=0.49]{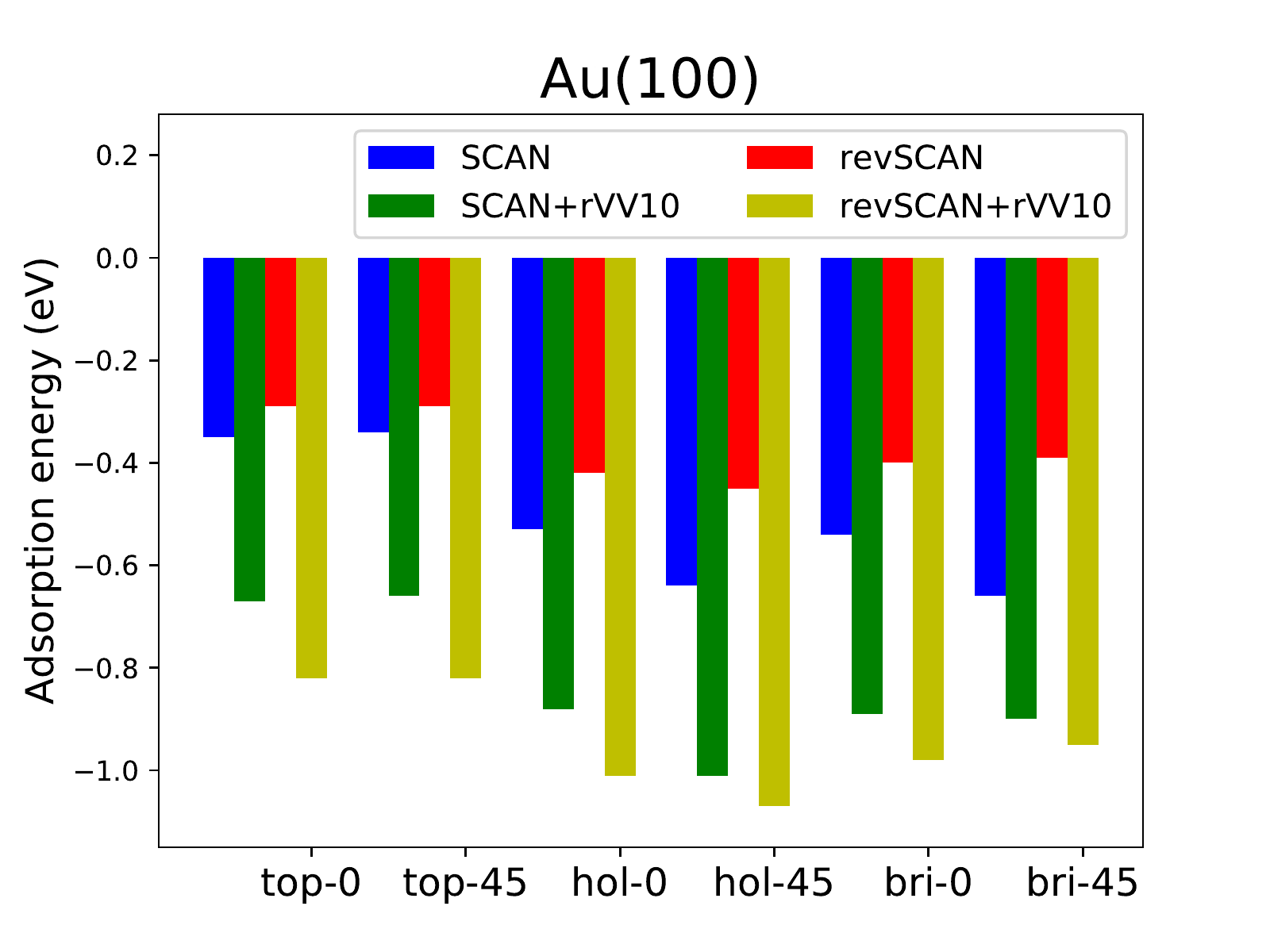} 
		\includegraphics[scale=0.49]{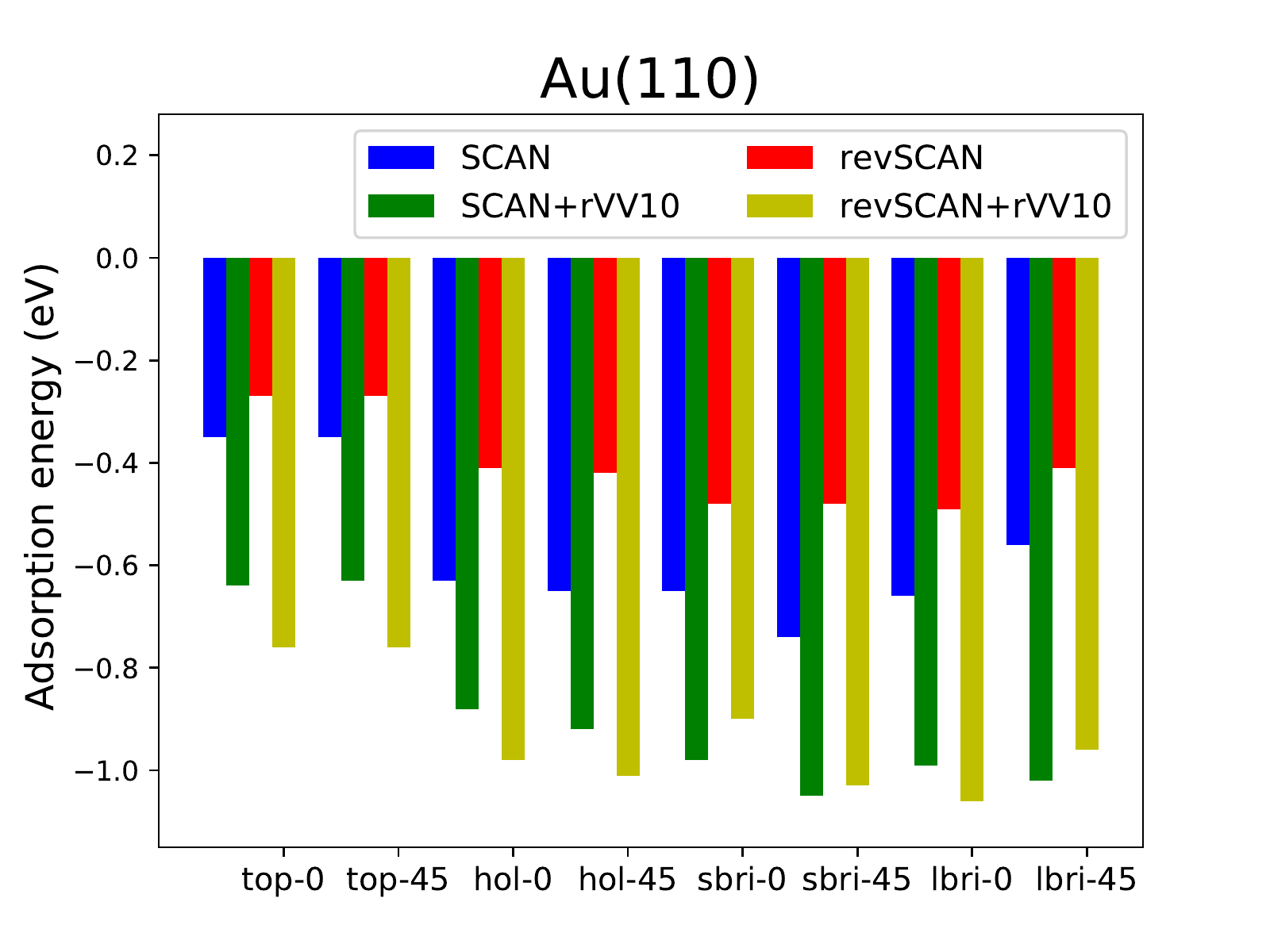} 
		\caption{(on left) Compares the site dependence of thiophene over Au(100) surface. Hollow-45 site as in the case of copper and silver is the most preferred site of adsorption. (on right) Compares the site dependence of thiophene on Au(110) surface. All the XC functionals except revSCAN predicts shortbridge-45 as the most stable site of adsorption. revSCAN predicts longbridge-0 as the most stable site in this system too.  }
	\end{figure}	
	\begin{table}[h!]
		\centering
		\renewcommand\thetable{S5}
		\resizebox{\columnwidth}{!}{%
			\begin{tabular}{rcccccccccr}
				\hline
				\multicolumn{1}{l}{Metal} & surface & \multicolumn{9}{c}{Functionals} \\
				\hline
				&       & PBE   & PBEsol & SCAN  & revSCAN & PBE+rVV10 & PBEsol+rVV10 & SCAN+rVV10 & revSCAN+rVV10 & \multicolumn{1}{c}{Reference$\cite{tyson1975,tyson1977}$} \\
				\hline
				& 111   & 1.28  & 1.60  & 1.56  & 1.58  & 1.61  & 1.95  & 1.77  & 1.91  &  \\
				\multicolumn{1}{l}{Cu} & 110   & 1.51  & 1.87  & 1.83  & 1.84  & 1.84  & 2.23  & 2.04  & 2.19  & \multicolumn{1}{c}{1.79 $\pm$ 0.19} \\
				& 100   & 1.44  & 1.79  & 1.75  & 1.75  & 1.77  & 2.15  & 1.97  & 2.10  &  \\
				&       &       &       &       &       &       &       &       &       &  \\
				& 111   & 0.69  & 0.97  & 0.91  & 0.91  & 1.04  & 1.35  & 1.13  & 1.28  &  \\
				\multicolumn{1}{l}{Au} & 110   & 0.90  & 1.24  & 1.15  & 1.15  & 1.26  & 1.63  & 1.39  & 1.54  & \multicolumn{1}{c}{1.51 $\pm$ 0.16} \\
				& 100   & 0.87  & 1.18  & 1.11  & 1.10  & 1.22  & 1.57  & 1.34  & 1.47  &  \\
				&       &       &       &       &       &       &       &       &       &  \\
				& 111   & 0.73  & 0.99  & 0.95  & 0.96  & 1.03  & 1.33  & 1.15  & 1.28  &  \\
				\multicolumn{1}{l}{Ag} & 110   & 0.86  & 1.16  & 1.11  & 1.11  & 1.17  & 1.51  & 1.31  & 1.44  & \multicolumn{1}{c}{1.25 $\pm$ 0.13} \\
				& 100   & 0.81  & 1.10  & 1.06  & 1.05  & 1.12  & 1.44  & 1.25  & 1.36  &  \\
				\hline
			\end{tabular}}
			\label{tab:addlabel}%
			\caption{Surface energy(in $J/m^2$) of different surfaces of copper, gold and silver. The average of the surface energies of three different surfaces are compared against the reference$\cite{tyson1975,tyson1977}$ values in Table 4 in the main text. rVV10 corrections on meta-GGA's significantly improve the surface energies. Both SCAN+rVV10 and revSCAN+rVV10 overall perform better for surface energies. }
		\end{table}%
		\pagebreak
		\textbf{\Large{Redhead's peak maximum method:}}\\
		Redhead's peak maximum method (Readhead's Analysis)$\cite{de1990thermal,redhead1962thermal}$ has been utilized to estimate the adsorption energy of the thiophene over Cu(111), Au(111) and Ag(111) surfaces from the thermal desorption spectra (TDS) . With the information of the peak maximum temperature ($T_m$), adsorption energy ($E_{ad}$) is calculated as:
		\begin{equation}
		E_{ad}=-RT_{m}[ln(\frac{\nu T_{m}}{\beta})-3.46]
		\end{equation} 
		Where R is the ideal gas constant, $\nu$ is the pre exponential factor of desorption (commonly called prefactor) and $\beta$ is the heating rate. The value of $\nu$ is chosen $\cite{de1990thermal}$ as $10^{13} s^{-1}$. The value of $T_m$ and $\beta$ are utilized from the TDS (TPD) experiments.$\cite{milligan2001complete,liu2002chemistry,vaterlein2000orientation}$ Notice that the accuracy of Redhead's method was questioned, but our analysis shows that despite its simplicity it gives reasonable adsorption energies for benzene and thiophene on the coinage metal surfaces. For example it was stated that "Temperature-programed desorption (TPD) studies of bezene/Ag(111) yield a desorption temperature of 220 K. Depending on the value of the empirical prefactor in the Redhead equation utilized to convert this temperature into  EAds, values ranging from 0.43 eV to 0.80 eV have been reported in the literature [16–19]."$\cite{liu2015quantitative}$  Substitution $T_m$ = 220 K and $\beta$ = 1 K/s yields EAds = -0.60 eV (not -0.43 eV or -0.8 eV). Applying the more correct 230 K, the vanishing surface coverage limit, gives -0.63 eV. This value is within the error bar of the result of the complete analysis: -0.68 $\pm$ 0.05eV for Benzene on Ag(111) in the limit of vanishing surface coverage. It seems that $\nu$ =  $10^{13} s^{-1}$ might yield reasonable values. The claim that "Coverage dependence of the pre-exponential factor can introduce uncertainties of up to 0.2 eV for the molecules considered here."$\cite{christian2016surface}$ is not supported by our results if the lowest published coverage leading to highest $T_m$ is considered and $\nu$ =  $10^{13} s^{-1}$ in the Redhead's adsorption energy estimation. The precision of the Redhead's method is around or better than 0.1 eV for thiophene or benzene adsorption on coinage metals. Notice also that deviations of eq. (1) from the analytically correct expression are within 1.5\% provided that $\nu$/$\beta$ falls between $10^{8}$ and $10^{13} K^{-1}$ for first order desorption.
		\begin{table}[h!]
			\centering
			\renewcommand\thetable{S6}
			\begin{tabular}{lccc}
				\hline
				& $T_m$    & $\beta$  & Adsorption energy \\
				\hline
				Surface & (K)   & (K/s) & (eV) \\
				\hline
				Cu(111) & 234$\cite{milligan2001complete}$   & 0.5   & -0.66 \\
				Au(111) & 240$\cite{liu2002chemistry}$   & 3     & -0.64 \\
				Ag(111) & 190$\cite{vaterlein2000orientation}$   & 1     & -0.52 \\
				\hline
			\end{tabular}%
			\label{tab:addlabel}%
			\caption{Calculation of adsorption energies of thiophene over Cu(111), Au(111) and Ag(111) using Redhead's Analysis. With the information of peak maximum temperature ($T_m$) and the heating rate ($\beta$) utilized in the experiment's$\cite{milligan2001complete,liu2002chemistry,vaterlein2000orientation}$ such energies can be estimated.}
		\end{table}%

\providecommand{\latin}[1]{#1}
\makeatletter
\providecommand{\doi}
{\begingroup\let\do\@makeother\dospecials
	\catcode`\{=1 \catcode`\}=2 \doi@aux}
\providecommand{\doi@aux}[1]{\endgroup\texttt{#1}}
\makeatother
\providecommand*\mcitethebibliography{\thebibliography}
\csname @ifundefined\endcsname{endmcitethebibliography}
{\let\endmcitethebibliography\endthebibliography}{}



\end{document}